\pdfoutput=1
\documentclass[
reprint,
superscriptaddress,
showpacs,
amsmath,
amssymb,
aps,
pra,
longbibliography,
]{revtex4-1}

\usepackage{graphicx}
\usepackage{dcolumn}
\usepackage{bm}
\usepackage{color}
\usepackage{url}
\usepackage{braket}
\usepackage{subcaption}

\usepackage[colorlinks=true,urlcolor=blue]{hyperref}

\usepackage[english]{babel}
\usepackage{letltxmacro}

\LetLtxMacro{\ORIGselectlanguage}{\selectlanguage}
\makeatletter
\DeclareRobustCommand{\selectlanguage}[1]{%
  \@ifundefined{alias@\string#1}
    {\ORIGselectlanguage{#1}}
    {\begingroup\edef\x{\endgroup
       \noexpand\ORIGselectlanguage{\@nameuse{alias@#1}}}\x}%
}
\newcommand{\definelanguagealias}[2]{%
  \@namedef{alias@#1}{#2}%
}
\makeatother

\definelanguagealias{en}{english}

\newcommand{\figref}[1]{Fig.~\ref{#1}}

\begin{document}
\title{Fault-tolerance thresholds for the surface code with fabrication errors}

\author{James M.~Auger}
\email{james.auger.09@ucl.ac.uk}
\affiliation{
 Department of Physics and Astronomy,
 University College London,
 Gower Street,
 London,
 WC1E 6BT,
 UK
}
\author{Hussain Anwar}
\affiliation{
 Department of Physics,
 Imperial College London,
 London,
 SW7 2AZ,
 UK
}
\affiliation{
 Department of Physics and Astronomy,
 University College London,
 Gower Street,
 London,
 WC1E 6BT,
 UK
}
\author{Mercedes Gimeno-Segovia}
\affiliation{Quantum Engineering Technology Labs, H. H. Wills Physics Laboratory and Department of Electrical and Electronic Engineering, University of Bristol, BS8 1FD, UK}
\affiliation{Institute for Quantum Science and Technology, University of Calgary, Alberta T2N 1N4, Canada}
\affiliation{
 Department of Physics,
 Imperial College London,
 London,
 SW7 2AZ,
 UK
}
\author{{Thomas M.~Stace}}
\affiliation{
 ARC Centre for Engineered Quantum Systems,
 University of Queensland,
 Brisbane 4072,
 Australia
}

\author{Dan E.~Browne}
\affiliation{
 Department of Physics and Astronomy,
 University College London,
 Gower Street,
 London,
 WC1E 6BT,
 UK
}

\begin{abstract}
The construction of topological error correction codes requires the ability to fabricate a lattice of physical qubits embedded on a manifold with a non-trivial topology such that the quantum information is encoded in the global degrees of freedom (i.e. the topology) of the manifold. However, the manufacturing of large-scale topological devices will undoubtedly suffer from fabrication errors---permanent faulty components such as missing physical qubits or failed entangling gates---introducing permanent defects into the topology of the lattice and hence significantly reducing the distance of the code and the quality of the encoded logical qubits.
In this work we investigate how fabrication errors affect the performance of topological codes, using the surface code as the testbed. 
A known approach to mitigate defective lattices involves the use of primitive SWAP gates in a long sequence of syndrome extraction circuits. Instead, we show that in the presence of fabrication errors the syndrome can be determined using the supercheck operator approach and the outcome of the defective gauge stabilizer generators without any additional computational overhead or the use of SWAP gates. We report numerical fault-tolerance thresholds in the presence of both qubit fabrication and gate fabrication errors using a circuit-based noise model and the minimum-weight perfect matching decoder.
Our numerical analysis is most applicable to 2D chip-based technologies, but the techniques presented here can be readily extended to other topological architectures. We find that in the presence of $8\%$ qubit fabrication errors, the surface code can still tolerate a computational error rate of up to $0.1\%$.

\end{abstract}
\maketitle
 
\section{Overview}

Implementing topological quantum error correction codes has become the focus of many current experiments, with recent advances made in building small codes consisting of small numbers of physical qubits with fine-tuned local interactions \cite{qecexp1,qecexp2,qecexp3,qecexp4,qecexp5,qecexp6}, hence paving the way for potential scalability in the coming years. However, it is expected that a universal topological fault-tolerant architecture will have an enormous number of physical components, and for such a large-scale machine the manufacturing and fine-tuning of each individual component will undoubtedly suffer from permanent faults resulting from imperfect manufacturing processes---we refer to such faults as \emph{fabrication errors}.  As an example, the latest non-universal D-Wave 2X machine has qubit manufacturing defects (typically fewer than 5\%)~\cite{K15baq}.  Therefore, it is important that the performance of the current schemes is studied against such a static error.

\begin{figure}
 \includegraphics[width=0.8\columnwidth]{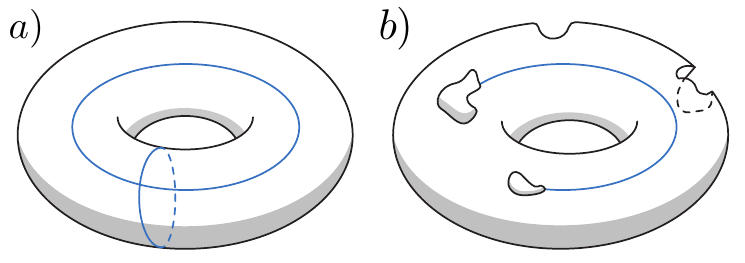}
 \caption{\label{fig:torus} a) The construction of topological codes (depicted here using the toy example of the toric code) relies on the fabrication of a perfect topology in order for the encoded logical state to be globally protected. b) Fabrication errors have the drastic effect of damaging the topology and shortening the distance of the code.}
\end{figure}

Surface codes are a family of stabilizer quantum error correction codes defined on a two-dimensional manifold with local vertex and plaquette stabilizer generators; the prototypical example is the toric code, introduced originally by Kitaev~\cite{K03ftq}, see ~\figref{fig:torus}a. This work focuses on using the square planar surface code~\cite{B98qco} as a memory for a single logical qubit, but the close similarity between different types of stabilizer topological codes mean the results and techniques presented here are applicable to other families of codes, such as color codes \cite{BMD06tqd}. 

The construction of topological codes relies on utilizing the non-trivial topology of a lattice to encode a logical state, such that the encoded state can only be corrupted by global errors. Such a construction is susceptible to fabrication errors by design, as fabrication errors directly damage the topology and introduce new degrees of freedom as well as lowering the code distance for the encoded logical states, see~\figref{fig:torus}b.

The threshold performance of the surface codes has been extensively studied against many types of noise models, such as for Pauli errors~\cite{Wang2003-mh,WFH11scq,S14ftt}, stochastic qubit loss~\cite{SBD09tft,Stace2010-qc}, leakage~\cite{SCG15lsi,Fowler2013-bc,Whiteside2014-ji,GF15lra,WG17qac}, and general non-Pauli local noise for small codes~\cite{TK14llds,DP16tns}, but much less focus has been given to errors resulting from imperfect manufacturing processes.
In this paper we investigate the threshold performance of the surface code in the presence of fabrication errors, and we  show that the ability to disable a qubit or an entangling gate (a link) is sufficient to map any fabrication error into disabled qubits, hence allowing us to always form larger defected stabilizer operators, the so-called \emph{supercheck operators}~\cite{SBD09tft}. The scheme we present does not require anything that is not already necessary to perform computation by code deformation~\cite{R07ftq, FMM12sct} or lattice surgery~\cite{H12scq}. We will not discuss specific specific experimental implementations, but the techniques presented here could be directly applicable to chip-based designed for topological computation.

Two recent approaches have been proposed to mitigate the effect of fabrication errors; the first is to construct a robust topology that tolerates sparse fabrication errors using additional local graph encoding~\cite{YG16r}, and the second is to use primitive $\mathrm{SWAP}$ gates in the construction of the syndrome read-out circuit~\cite{NFH16sce}. Our approach differs from these by keeping the construction of the surface code unaltered without attempting to mitigate the missing components or performing an adaptive procedure. We show that by performing (defected) gauge stabilizer operators \cite{B10tsc,BD13ssc}, supercheck operator outcomes can be obtained deterministically, see Sec.~\ref{sec:fabrication}. Our main result is summarised in \figref{fig:fabrication_thresholds}, which shows Pauli error thresholds obtained by simulations of the surface code in the presence of fabrication errors.

This paper is structured as follows. In Sec.~\ref{sec:surface_code} we introduce the components of the surface code and the syndrome extraction circuits. In Sec.~\ref{sec:fabrication} we define the types of fabrication errors and how to perform the syndrome extraction circuits on a defected lattice by constructing the outcome of supercheck operators from the gauge qubit operators. We outline our noise model and Monte-Carlo simulation in Sec.~\ref{sec:simulation} and present the fault-tolerant thresholds we obtained in Sec.~\ref{sec:thresholds}. In the final two sections we discuss our approach in comparison to other schemes and conclude.

\section{The Surface Code}\label{sec:surface_code}

\begin{figure}
 \includegraphics[width=.8\columnwidth]{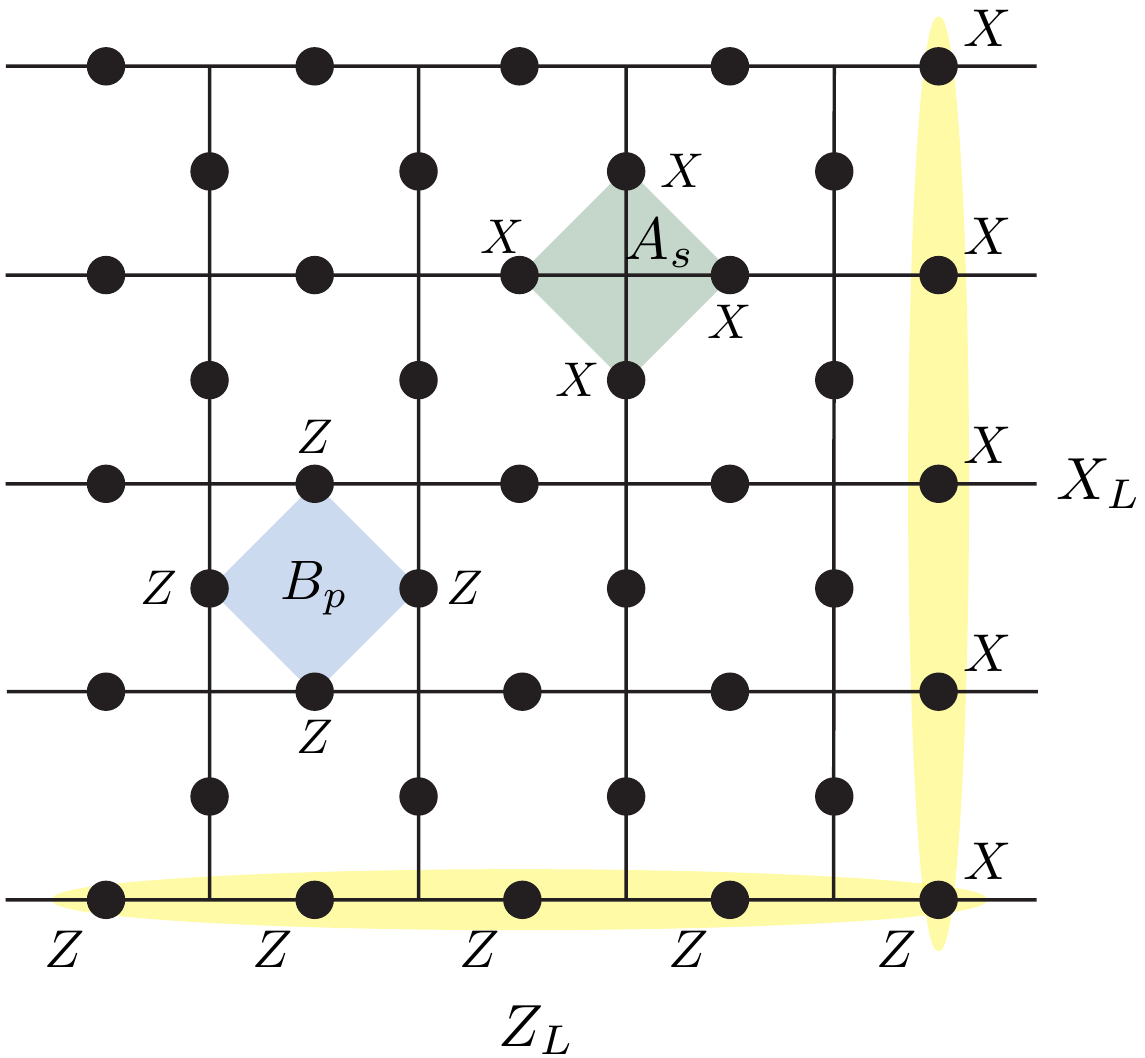}
 \caption{\label{fig:surface_code}  An example of a primal lattice of a distance $5$ surface code. Each star $s$ and plaquette $p$ of the lattice is associated with a four-body $X$-type ($A_s$) and $Z$-type ($B_p$) stabilizer generator respectively. The logical operators $X_{L}$ and $Z_{L}$ correspond to anti-commuting strings of operators that span the lattice but each commute with all the stabilizer generators. The left and right edges are the \emph{rough} edges and the top and bottom edges are the \emph{smooth} edges.}
\end{figure}

We define the surface code on an $L\times L$ square lattice with open boundaries as shown in~\figref{fig:surface_code}, such that each edge represents a physical data qubit. The stabilizers of the code are generated by $X$-type and $Z$-type four-body operators associated with each star (or vertex) and each plaquette of the lattice, denoted here by $A_s$ and $B_p$, respectively. Collectively, the star and the plaquette operators are referred to as the \emph{check operators}. The logical Pauli $X_L$ and $Z_L$ operators of the code are global anti-commuting string-like operators that span the lattice and commute with all the stabilizer generators, examples are shown on the primal lattice in~\figref{fig:surface_code}. In other words, the logical $Z_L$ ($X_L$) operators have support over non-contractible loops on the primal (dual) lattice.

The action of a logical operator on a logical state of the surface code is invariant under a multiplication by a stabilizer generator, so many equivalent logical operators can be defined. The distance of the code corresponds to the weight of the shortest possible logical operator. Therefore, a perfectly fabricated surface code will have a code distance $L$. Increasing the lattice size would offer better protection against errors as the number of individual errors required to cause a logical error is greater.

Error detection proceeds by measuring all the stabilizer generators \cite{DKL02tqm}, such that when there are no errors, the outcome of all measurements is $+1$. Measuring all stabilizer generators collapses arbitrary qubit errors into Pauli $X$ and $Z$ errors on the data qubits~\cite{Nielsen2000-nv, DKL02tqm}; if an error anti-commutes with a stabilizer generator then the outcome of that stabilizer generator is flipped to $-1$. If a string of a single-type of Pauli error occurs, only the ends of the string are detectable, and different error strings may lead to the same detection pattern. The set of outcomes of stabilizer generator measurements is called the \emph{syndrome}.

An error does not have to be corrected by exactly inverting the Pauli errors that caused it. Any string of errors and corrections that is a product of stabilizer generators is harmless (forming a contractible loop), however a string of errors and corrections that is equivalent to a logical operation leads to a failure. After obtaining the syndrome measurements, their values are passed to a \emph{decoder}---a classical algorithm that finds the correction that is least-likely to result in a logical error (or at least a good approximation to this correction). In our analysis, we use the well-studied minimum-weight perfect-matching (MWPM) algorithm for our decoder~\cite{DKL02tqm}.

\begin{figure}
 \includegraphics[width=\columnwidth]{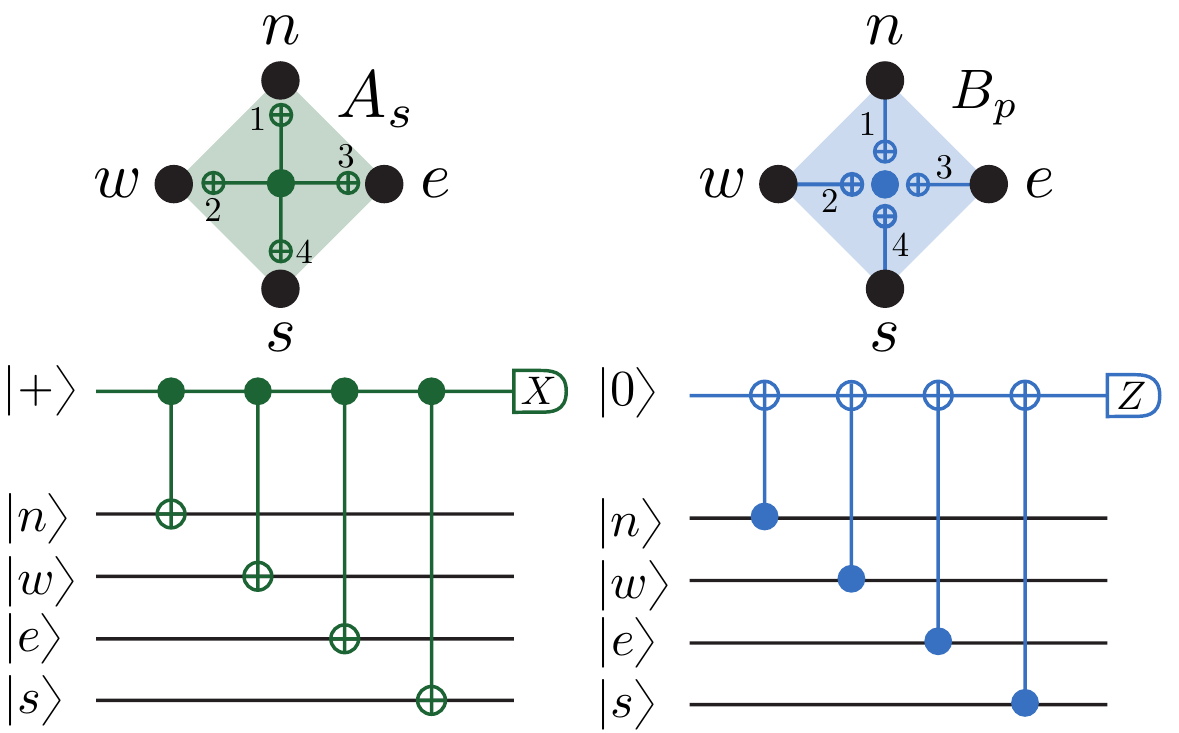}
 \caption{\label{fig:order_of_measurement}Syndrome extraction circuits for $X$-type (left) and $Z$-type (right) stabilizer generators. Each syndrome extraction circuit involves six temporal steps: preparation of the syndrome qubit, four CNOT gates, and syndrome qubit measurement. The data qubits are idle during the preparation and measurement steps. The CNOT gates are always applied in a specific order, in this case north (\textit{n}), west (\textit{w}), east (\textit{e}) then south (\textit{s}) (forming a zigzag shape), in order to ensure all stabilizer generators commute when measured.}
\end{figure}

The stabilizer generators are multi-qubit operators, which can be difficult to implement experimentally. To assist in measuring each stabilizer generator, an ancilla qubit, often called the \emph{syndrome qubit}, is associated with each star and each plaquette operator in a local \emph{syndrome-extraction circuit}, as shown in~\figref{fig:order_of_measurement}. The syndrome qubit is entangled to the qubits that support the stabilizer by an ordered sequence of two-qubit gates. Once this entangling procedure is executed, the outcome of the stabilizer generator is determined by measuring the ancilla qubit in the Pauli basis. By performing the two-qubit gates in a \emph{z-shaped} order, as shown in~\figref{fig:order_of_measurement}, all the stabilizer generators can be measured simultaneously in a total of six time-steps: one for syndrome qubit preparation, four for two-qubit gates and one for syndrome qubit measurement (it may be possible to combine measurement and initialization for some implementations, but we consider them to be separate processes in our simulations). These six time steps constitute one round of syndrome measurement.

The components of the syndrome extraction circuit are themselves prone to the same type of errors as the data qubits of the code, which could lead to an incorrect (false) error detection event. Therefore, multiple rounds of syndrome measurements, $O(L)$, are needed to improve the confidence in the syndrome outcomes. The repeated rounds of syndrome measurements creates a 3D syndrome history, which can be processed by the decoder to return  a correction operator on the physical lattice.

\section{Fabrication Errors}\label{sec:fabrication}

We define fabrication errors as permanent faulty components caused during the initial chip-manufacturing process of the surface code, or due to failed components arising during the lifetime of the chip. It is important to emphasize that the location of the fabrication errors is known before the surface code is used in the computation (i.e. the fabrication errors are known deterministic missing components). In addition, we assume that the user of the surface code chip is able to turn-off (or stop) any of the components.  We consider two types of fabrication error: \emph{qubit fabrication} errors and \emph{link fabrication} errors. A qubit fabrication error is considered to be a qubit (either a data qubit or syndrome qubit) that is permanently faulty and cannot be used to store quantum information reliably. A link fabrication error is considered to be an error that prevents two qubits from interacting, i.e. it prevents a CNOT or CPHASE gate from being performed. 

Before proceeding, it is useful to introduce some additional terminology to describe the different types of failed components we will encounter in our analysis. We use the term \emph{faulty} to strictly refer to a component with a permanent fabrication error, and the term \emph{disabled} to refer to a component that we have chosen to disable. Moreover, we call a check operator \emph{damaged} if at least one  of its four links suffers a fabrication error.

Fabrication errors are detrimental for the surface code construction; left unchecked, they can introduce new degrees of freedom (i.e. extra logical qubits). For example, if a syndrome qubit is faulty, one might be tempted to simply disable the associated stabilizer generator. But a disabled star or plaquette creates a new logical qubit that can interact with our desired logical qubit, therefore reducing the code distance. A reduction in code distance alone may not be a problem in itself, but if we assume fabrication errors are randomly spaced throughout the code then code distance will start to shrink with increasing $L$, leading to only a pseudo-threshold behavior for smaller lattice sizes that disappears for larger lattices.
 
We will now show how the detrimental effect of qubit and link fabrication errors can be mitigated at no additional hardware cost by disabling data qubits.

\subsection{Measuring supercheck operators and gauge qubits}

The idea of using supercheck operators was first introduced to combat lost data qubits in the toric code~\cite{SBD09tft}. This approach works on the basis that the product of two stabilizer generators is also in the code stabilizer, so when a data qubit is lost (i.e. an edge is removed from the lattice) the two adjacent damaged stabilizer generators can be jointly measured---forming a larger \emph{supercheck} operator---to avoid the lost data qubit, hence preserving the stabilizer structure of the code.  

\begin{figure}
  \includegraphics[width=\columnwidth]{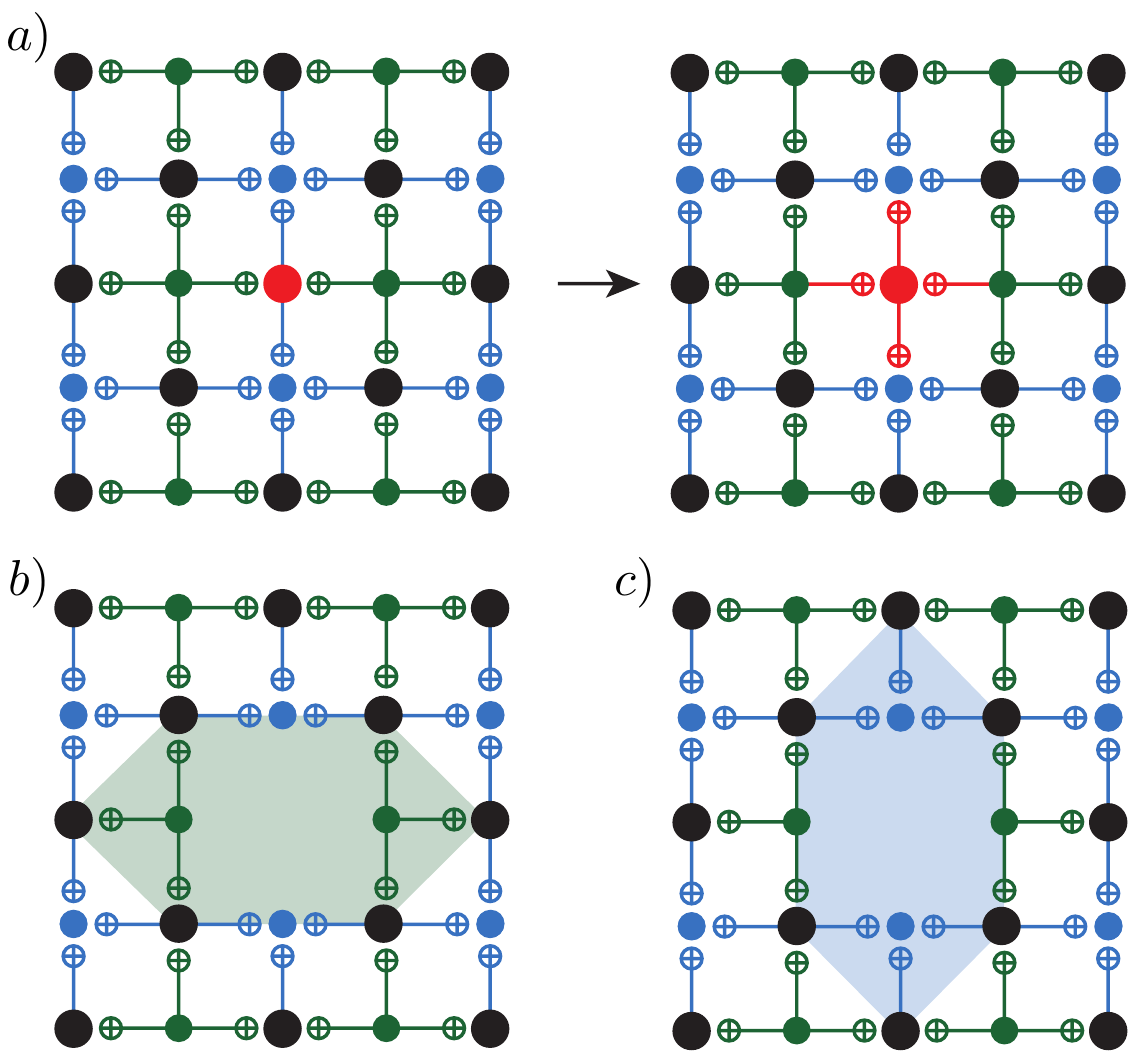}
  \caption{\label{fig:superchecks} When a data qubit is disabled or faulty (shown in red), the associated links are disabled (a), resulting in four adjacent damaged check operators. The product of two stabilizer generators is used to form a supercheck operator, effectively removing this qubit from the code.  This process occurs in both the primal (b) and dual (c) lattices; the CNOT gates shown in this figure are those used when measuring the supercheck operators as products of gauge operators.}
\end{figure}

In the context of fabrication errors, the same approach can be used for data qubit fabrication errors. However, measuring a supercheck operator directly is often non-trivial task as it may require interaction between arbitrarily separated qubits or involve many SWAP gates, as was shown in ~\cite{NFH16sce}, which can affect measurement of nearby stabilizer generators. Our approach for handling fabrication errors uses the same supercheck operator approach, but makes use of the concept of \emph{gauge qubits}. Instead of measuring the supercheck operators themselves, we use the gauge qubits to construct the outcome of the supercheck operator from the direct outcome of damaged stabilizer generators measurement, such that all interactions remain as nearest-neighbor qubit interactions and no SWAP gates are required; any additional processing required is performed classically.

Consider the simple case of disabling a data qubit as shown in \figref{fig:superchecks}; the adjacent damaged generators will anti-commute, but their supercheck operator product remains deterministic. Each disabled data qubit in the surface code (except data qubits on the edge of the lattice, see Sec.~\ref{sec:complications}) introduces one degree of freedom, or \emph{gauge} qubit (also called a ``junk qubit'' in \cite{SBD09tft,Stace2010-qc}), similar to when a stabilizer is turned-off to perform computation~\cite{FMM12sct}. The logical Pauli $X$ and $Z$ operators of these gauge qubits are the damaged star and plaquette stabilizers---we will refer to them as the \emph{gauge operators}. When these anti-commuting gauge operators are measured, the logical state of the gauge qubit is randomized, but the state of the gauge qubit is unimportant, so this randomization is not a problem. Importantly, strings of $X$ or $Z$ operators cannot terminate undetectably in this region, unlike when a stabilizer generator is turned off---these gauge operators reduce the code distance slightly, but code distance still scales with physical lattice size.

The supercheck operator product of damaged stabilizer generators commutes with every damaged stabilizer generator, so the supercheck operators remain in the stabilizer group and can be used for error correction during the classical processing stage by treating the products of the damaged star and plaquette as supercheck operators.

\subsection{Mapping fabrication errors to faulty data qubits}\label{sec:mapping}

We saw in the previous section how in the presence faulty (or disabled) data qubits the outcome of a supercheck operator can still deterministically be obtained by taking the product of the outcomes of the damaged (gauge) operators. We will exploit this fact to map both link fabrication errors and syndrome qubit fabrication errors to data qubit fabrication errors such that supercheck operators can always be formed. 

\begin{figure}
  \includegraphics[width=0.9\linewidth]{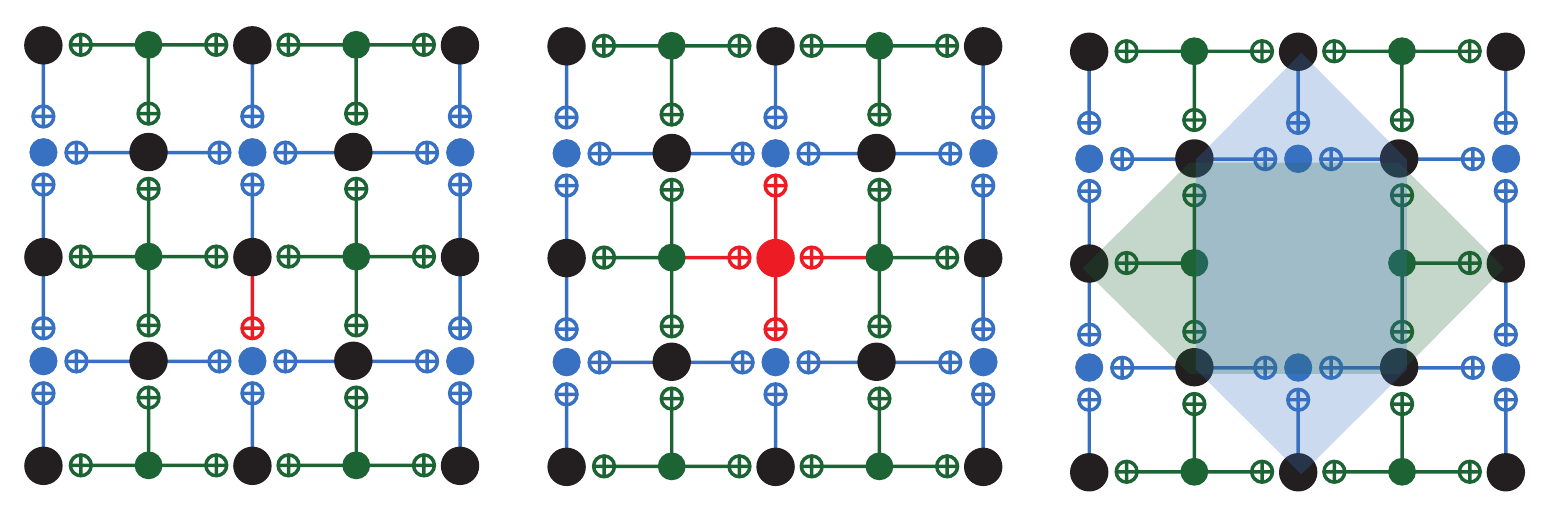}
  \caption{\label{fig:mapping_link}Mapping link fabrication errors to disabled data qubits; faulty and disabled components are shown in red. When a link fabrication error occurs (left), the data qubit associated with the link is disabled (middle) and superstars and superplaquettes are formed on the primal and dual lattices respectively (right).}
\end{figure}

\begin{figure}
  \includegraphics[width=0.9\linewidth]{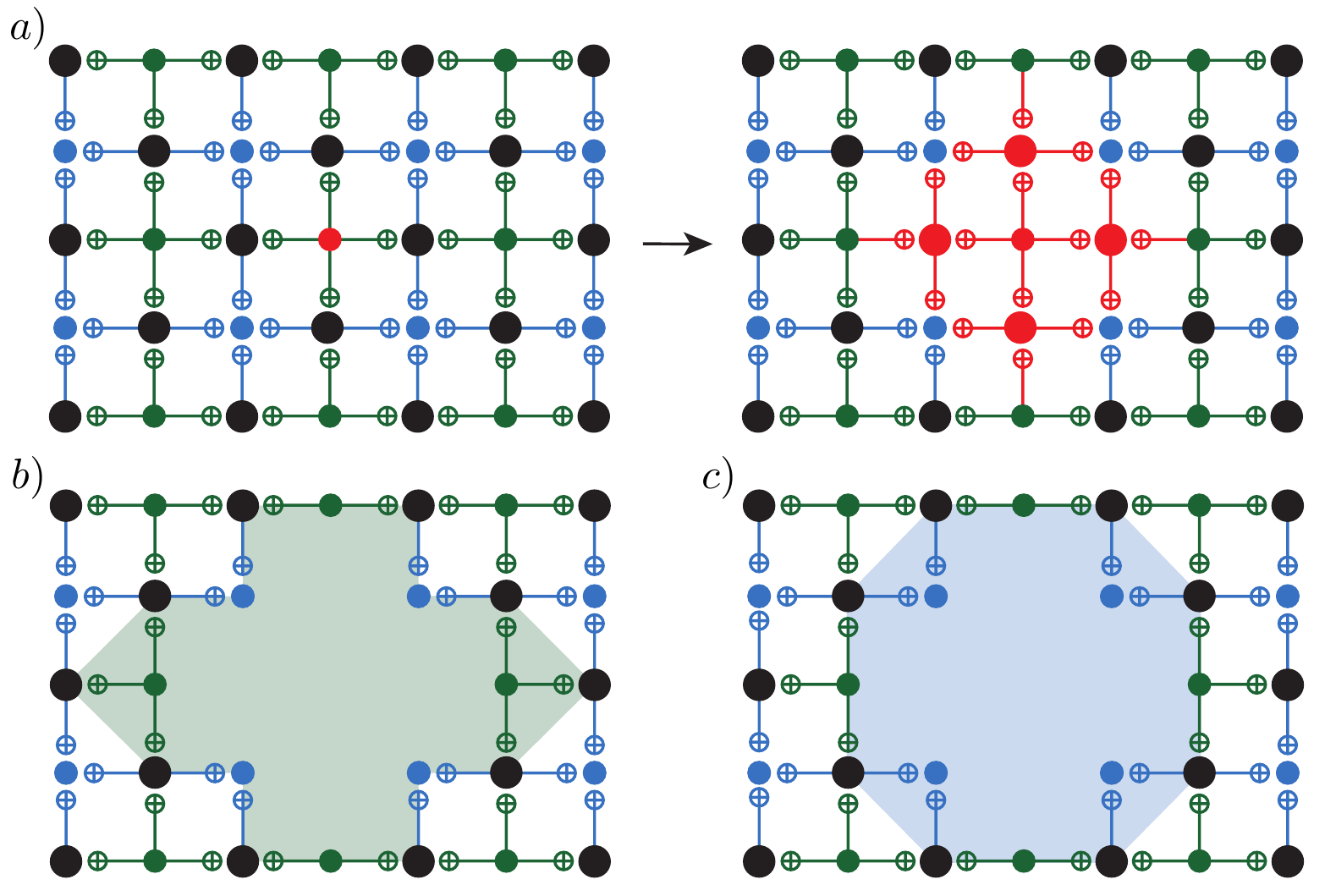}
  \caption{\label{fig:mapping_syndrome}Mapping syndrome fabrication errors to disabled data qubits; faulty and disabled components are shown in red. When a syndrome qubit fabrication error occurs, all the data qubits involved in the associated stabilizer generator are disabled (a). Large superstars (b) and superplaquettes (c) are formed around these disabled data qubits.}
\end{figure}

The mapping works as follows: 1) if a link fabrication error occurs, it is mapped to a qubit fabrication error on the data qubit to which it connects, so that the data qubit is disabled along with its associated links, as shown in~\figref{fig:mapping_link}. 2) If a syndrome qubit fabrication error occurs, it is mapped to qubit fabrication errors on all of the surrounding data qubits, so that all these data qubits along with their associated links are disabled, as shown in~\figref{fig:mapping_syndrome}. 

As a results we see that, in our approach, the syndrome qubit fabrication errors have the most destructive effect on the the lattice, which highlights an important bias between data and syndrome qubits. 


\subsection{Percolation thresholds and the effective code distance}
The use of supercheck operators is limited by percolation---if a string of faulty or disabled data qubits percolates the lattice, a logical qubit cannot be encoded as it is impossible to form consistent spanning logical operators and supercheck operators. Note that it would be possible to use part of the lattice as a smaller code, but we consider percolating fabrication defects to be a manufacturing failure as the intended distance of the code cannot be achieved by making a larger lattice, and hence the surface code is discarded. 

The square lattice structure of the surface code implies that the qubit percolation threshold for the surface code is equivalent to the bond percolation threshold for the square lattice, which is known analytic value of $50\%$. By using our mapping to data qubit fabrication errors, we can derive an approximate analytic percolation threshold in the presence of qubit and link fabrication errors using the following argument.

In the bulk of the lattice, each data qubit has four links. With a link fabrication error rate of $p_\mathrm{link}$, the probability of each data qubit being disabled due to faulty links is $1 - (1 - p_\mathrm{link})^4$, i.e. one minus the probability of no faulty links occurring. This implies that $50\%$ of data qubits will be disabled when $1 - (1 - p_\mathrm{link})^4 = 0.5$, or equivalently when $p_\mathrm{link} = 1 - \sqrt[4]{0.5} \approx 0.159$. This analysis does not account for qubits at the edges having fewer than four links, but the percolation threshold is an asymptotic behavior in the thermodynamic limit, so this effect can be neglected for large lattices.

An analogous argument can be used to calculate an approximate threshold for qubit fabrication errors. A data qubit will be disabled when it is either faulty or one or more of the four syndrome qubits it is linked to are faulty. The probability that a particular qubit is disabled when the qubit fabrication error rate is $p_\mathrm{qubit}$ is $1 - (1 - p_\mathrm{qubit})^5$. Therefore, the qubit fabrication error percolation threshold occurs approximately when $1 - (1 - p_\mathrm{qubit})^5 = 0.5$, or $p_\mathrm{qubit} = 1 - \sqrt[5]{0.5} \approx 0.129$. This analysis is less accurate than that for link fabrication errors as it does not take into account the possible correlations between disabled data qubits due to disabled syndrome qubits. However, localized clusters of disabled qubits are generally less likely to percolate than data qubits that are disabled at random, so we expect the actual threshold to be slightly higher.

\begin{figure}
	\includegraphics[width=\linewidth]{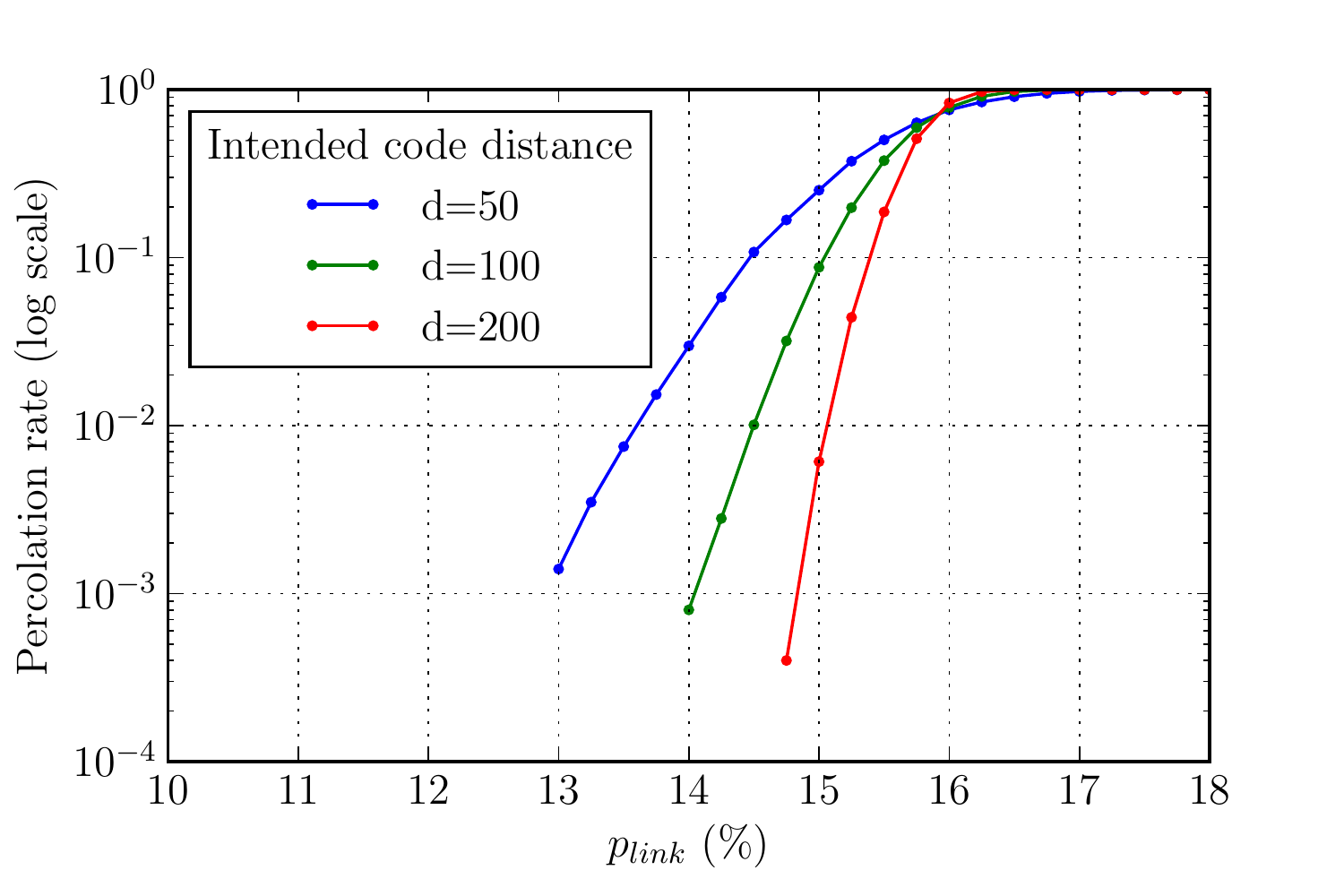}
    \caption{\label{fig:link_perc}Percolation rates for link fabrication errors only. The crossing point gives a threshold of just under 16\%.}
\end{figure}

\begin{figure}
	\includegraphics[width=\linewidth]{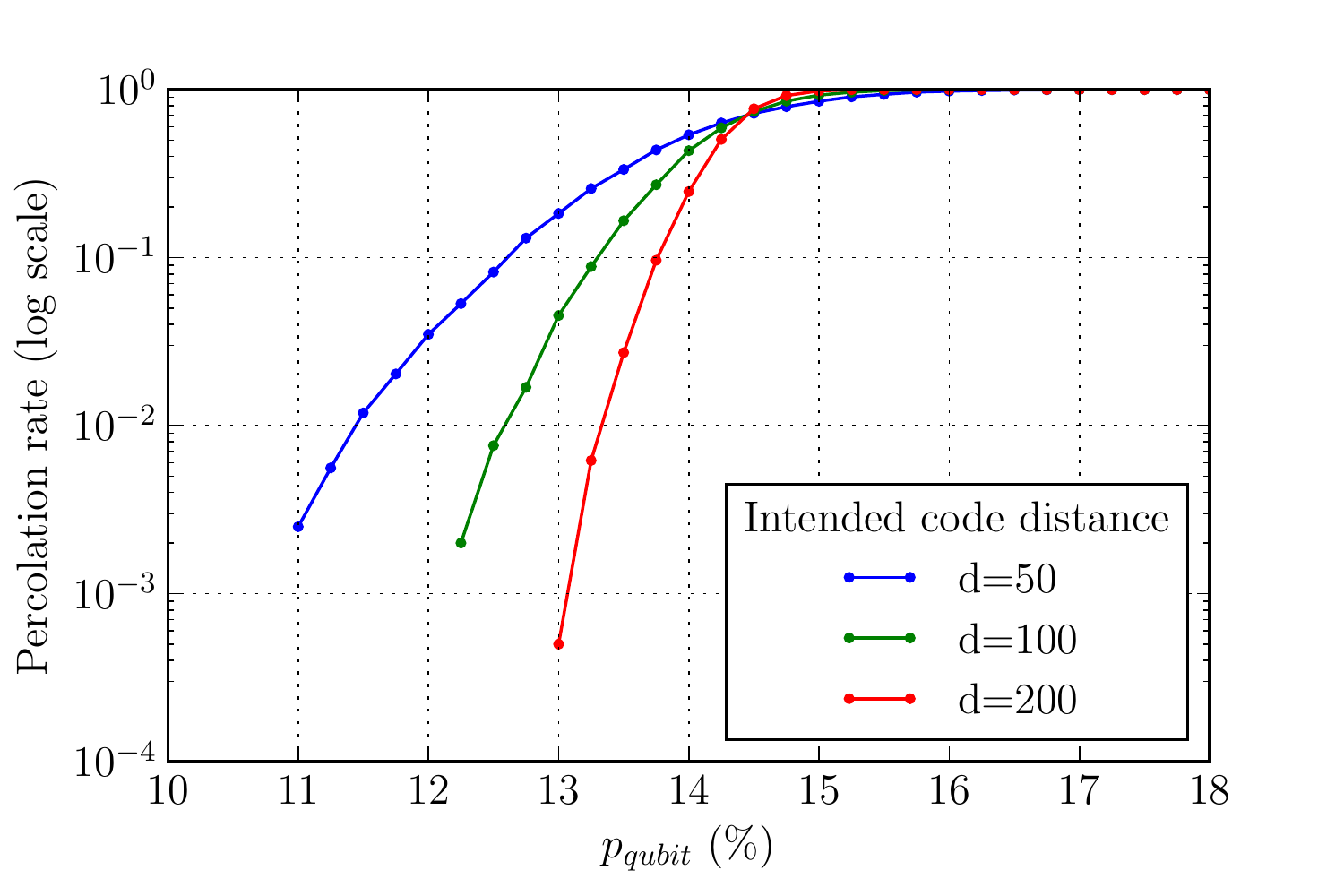}
    \caption{\label{fig:q_perc}Percolation rates for qubit fabrication errors only. The crossing point gives a threshold of around 14.5\%.}
\end{figure}

Our numerical simulations for percolation, shown in \figref{fig:link_perc} and \figref{fig:q_perc} for link and qubit fabrication errors, respectively, are in strong agreement with our above estimations. We find the link fabrication threshold, \figref{fig:link_perc}, to be just under $16\%$ and the qubit fabrication threshold, \figref{fig:q_perc}, to be between $14\%$ and $15\%$ (higher than $12.9\%$ due to data qubit loss occurring in localized clusters).

\begin{figure}
    \includegraphics[width=\linewidth]{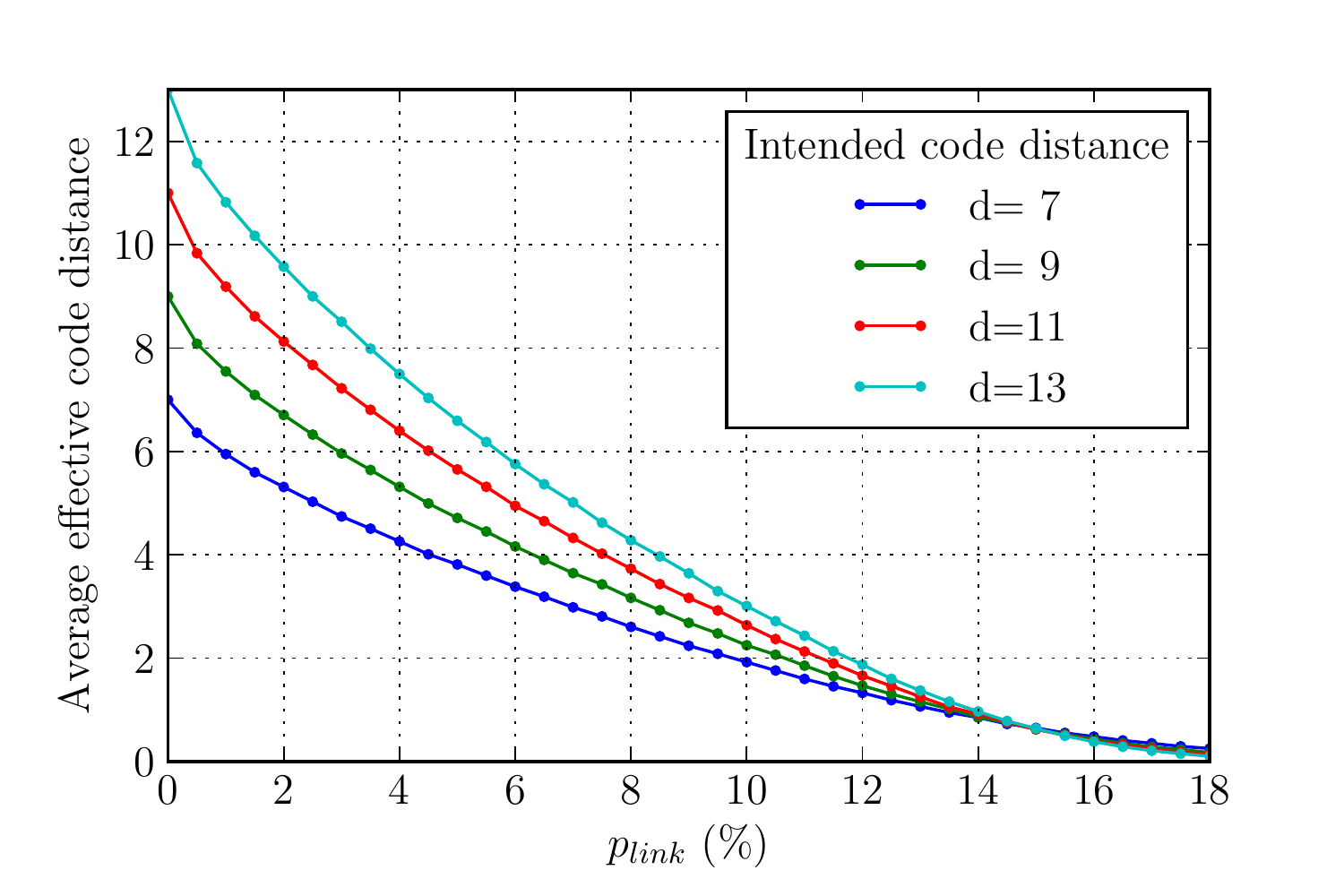}
    \caption{\label{fig:eff_dist_conn}Average effective code distance for link fabrication errors only.}
\end{figure}

\begin{figure}
    \includegraphics[width=\linewidth]{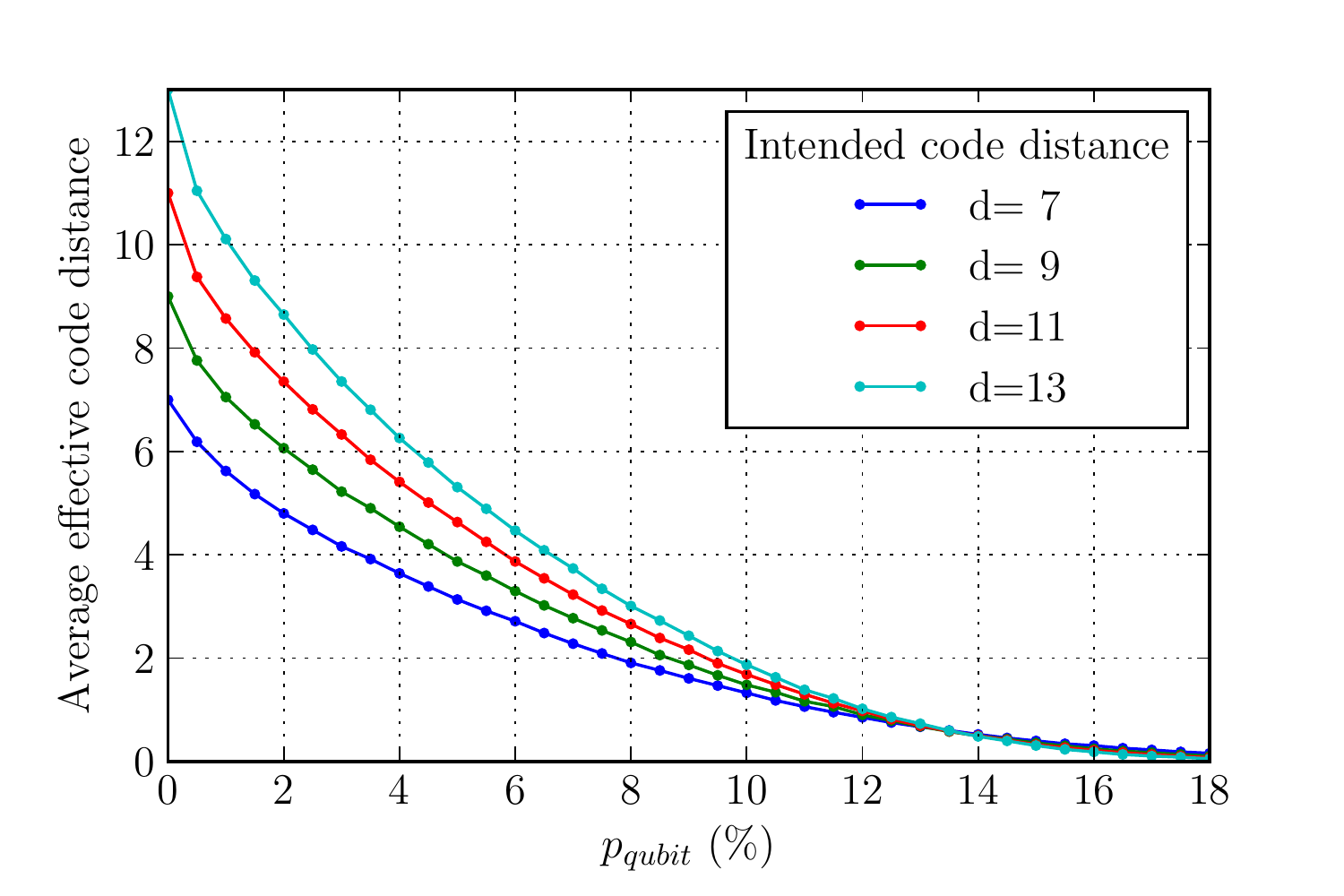}
    \caption{\label{fig:eff_dist_qubit}Average effective code distance for qubit fabrication errors only.}
\end{figure}

Forming supercheck operators leads to a reduction in the effective code distance compared to the intended code distance as it reduces the length of the shortest logical operator. We have analyzed how the average effective code distance varies with link and qubit fabrication errors, as shown in \figref{fig:eff_dist_conn} and \figref{fig:eff_dist_qubit}, respectively. These graphs were produced by simulating the fabrication of many planar code lattices with each type of fabrication error and then finding the effective code distance of each lattice by identifying the lowest-weight logical operator; the effective code distance was then averaged over all runs.


\subsection{Complications\label{sec:complications}}

There are two complications that occur when using gauge operators to measure supercheck operators. The first is that supercheck operators can only be measured during alternating rounds of syndrome measurement, and the second involves faulty data qubits at the edge of the lattice.

\begin{figure}
  \includegraphics[width=0.4\linewidth]{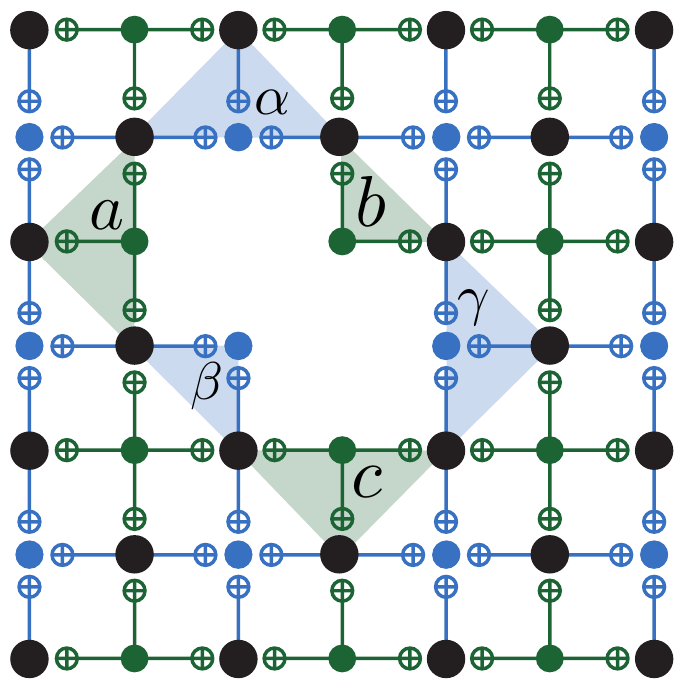}
  \caption{\label{fig:anticommutation}Effective order of gauge operator measurement. If the normal \textit{z-shaped} measurement pattern is used, the effective order in which the gauge operators is measured is $c$, $\gamma$ \& $\beta$, $a$ \& $b$, $\alpha$. The anticommutation randomizes the supercheck operator products, so $X$ and $Z$ type gauge operators are instead measured in alternating rounds to ensure the supercheck operator outcome is deterministic.}
\end{figure}

The interleaved \emph{z-shaped} measurement pattern that allows all stabilizer generators to be measured simultaneously on a perfect lattice can no longer be used for supercheck operators. Ensuring that the product of damaged stabilizers is deterministic requires that no anti-commuting operations are performed while the constituent gauge operators of the supercheck operators are being measured; this is not possible with the normal measurement pattern, as the example in~\figref{fig:anticommutation} shows. This problem is mitigated by measuring $X$-type and $Z$-type supercheck operators during alternating rounds. All undamaged stabilizer generators are measured every round as normal; this means that supercheck operators are measured half as frequently as undamaged stabilizer generators, and these measurements have a higher effective error rate than undamaged stabilizer generator measurements.

\begin{figure}
  \includegraphics[width=0.9\linewidth]{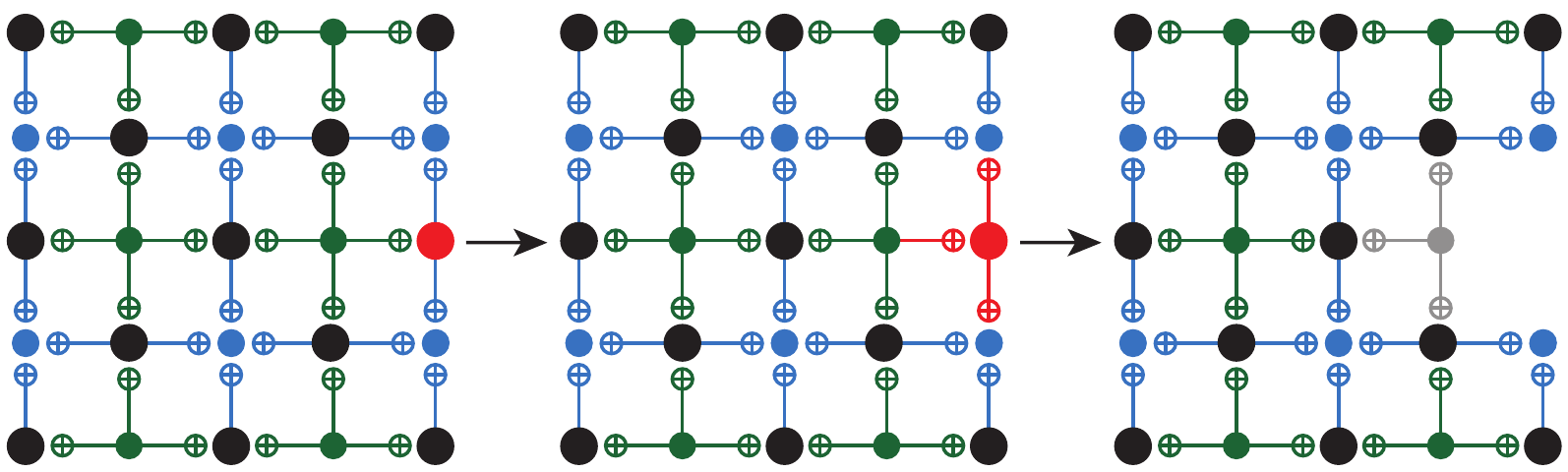}
  \caption{\label{fig:edge_problem}Dealing with fabrication errors at the edge of the lattice; faulty and disabled components are shown in red. When data qubits at the edge of the lattice are disabled (left, middle), the product of gauge operators is not in the stabilizer, so such stabilizer generators must be disabled rather than forming supercheck operators (right).}
\end{figure}

The second issue occurs when data qubits are faulty at the edges of the lattice. If there is a data qubit fault such as that shown in~\figref{fig:edge_problem}, then there is no corresponding stabilizer generator to pair it with. Therefore, the edge of the lattice must be redefined by completely disabling the damaged stabilizer generator. This process must then be repeated if any of the qubits in this new edge are faulty---the process effectively results in a supercheck operator being disabled.

\section{Simulation Methods}\label{sec:simulation}

\subsection{Error model}

Every gate is modelled to experience computational Pauli errors with a probability denoted here by the parameter $p_\mathrm{comp}$. Each two-qubit gate is assumed to act perfectly followed by a depolarizing Pauli noise with probability $p_\mathrm{comp}$. Single qubit gates (only the identity gate in our simulations) are assumed to act perfectly followed by depolarizing noise with probability $4p_\mathrm{comp}/5$. The justification for this follows that of~\cite{K05qcw}: $4p/5$ is the marginal error rate on each qubit involved in a two-qubit gate experiencing depolarizing noise with probability $p$. If we were to choose a single-qubit error rate of $p_\mathrm{comp}$, this would imply that single-qubit gates are more prone to errors than two-qubit gates, which is unlikely to be the case. Preparation is considered to have probability $p_\mathrm{comp}$ of preparing the state in an orthonormal basis, and measurement is considered to have a probability $p_\mathrm{comp}$ of giving the incorrect outcome.

All fabrication errors are considered to occur independently before error correction is initiated, and the locations of all fabrication errors are assumed to be known. A qubit fabrication error occurs with probability $p_\mathrm{qubit}$ for each qubit (syndrome and data qubits) and link fabrication errors occur with probability $p_\mathrm{link}$ for each link. The parameters $p_\mathrm{comp}$, $p_\mathrm{qubit}$ and $p_\mathrm{link}$ are varied independently.

\subsection{Simulations}

Each simulation starts by generating a lattice of qubits and links with the appropriate fabrication error rates. The fabrication errors are then mapped to data qubit fabrication using the mapping in Sec.~\ref{sec:mapping}.  Suitable logical operators are found by using a path finding routine on the primal and dual lattices. If a logical operator cannot be found, then the lattice is percolated by faulty (or disabled) data qubits and the simulation is aborted.

When a logical operator is found, $2\times L$ rounds of syndrome measurement are performed. Each round consists of syndrome qubit initialization, four stages of two-qubit gates and then syndrome qubit measurement. Each of these is considered to take one unit of time, and any qubit that is not involved in a two-qubit gate, measurement or preparation during a particular time step undergoes an identity gate.

The code is initialized with a round of perfect star and plaquette measurements to get the error-free outcomes for each star and plaquette measurement, and the simulations are capped with a final round of perfect measurement. All other rounds of stabilizer measurement use the Pauli error model given above. 

We use the CHP~\cite{AG04iso} stabilizer algorithm to simulate the quantum state of the lattice during all gates and measurements to ensure that the gauge operators give the correct outcomes. Once all measurements have been obtained, a MWPM routine involving Blossom~V~\cite{K09bva} is used to find a correction based on the measured syndrome. Edge weights for the perfect matching routine are set using the same methods as~\cite{BA13sor, SCG15lsi} to optimize the matching. Syndrome measurement is simulated on both the primal and dual lattices, but error correction is only performed on one lattice to reduce computation time (the symmetry between primal and dual lattices means that logical error rates are almost identical). Once the correction has been applied, we test for the occurrence of a logical $X$ error by checking if the combined error and correction string commutes with the logical $Z$ operator.

Pauli error rates are varied from $p_\mathrm{comp} = 0.05\%$ to $p_\mathrm{comp} = 1.00\%$ in steps of $0.05\%$, with additional values of $0.001\% < p_\mathrm{comp} < 0.010\%$ used when the Pauli error threshold becomes small. Each fabrication error rate is separately varied from from 0\% to 10\% in steps of 2\%, with an additional simulation performed at 5\% fabrication error rate to allow for a direct comparison with~\cite{NFH16sce}. Each combination of error rates and code distances is simulated for a minimum of $5 \times 10^4$ runs.

\section{Threshold results}\label{sec:thresholds}

\begin{figure}
  \includegraphics[width=1\linewidth]{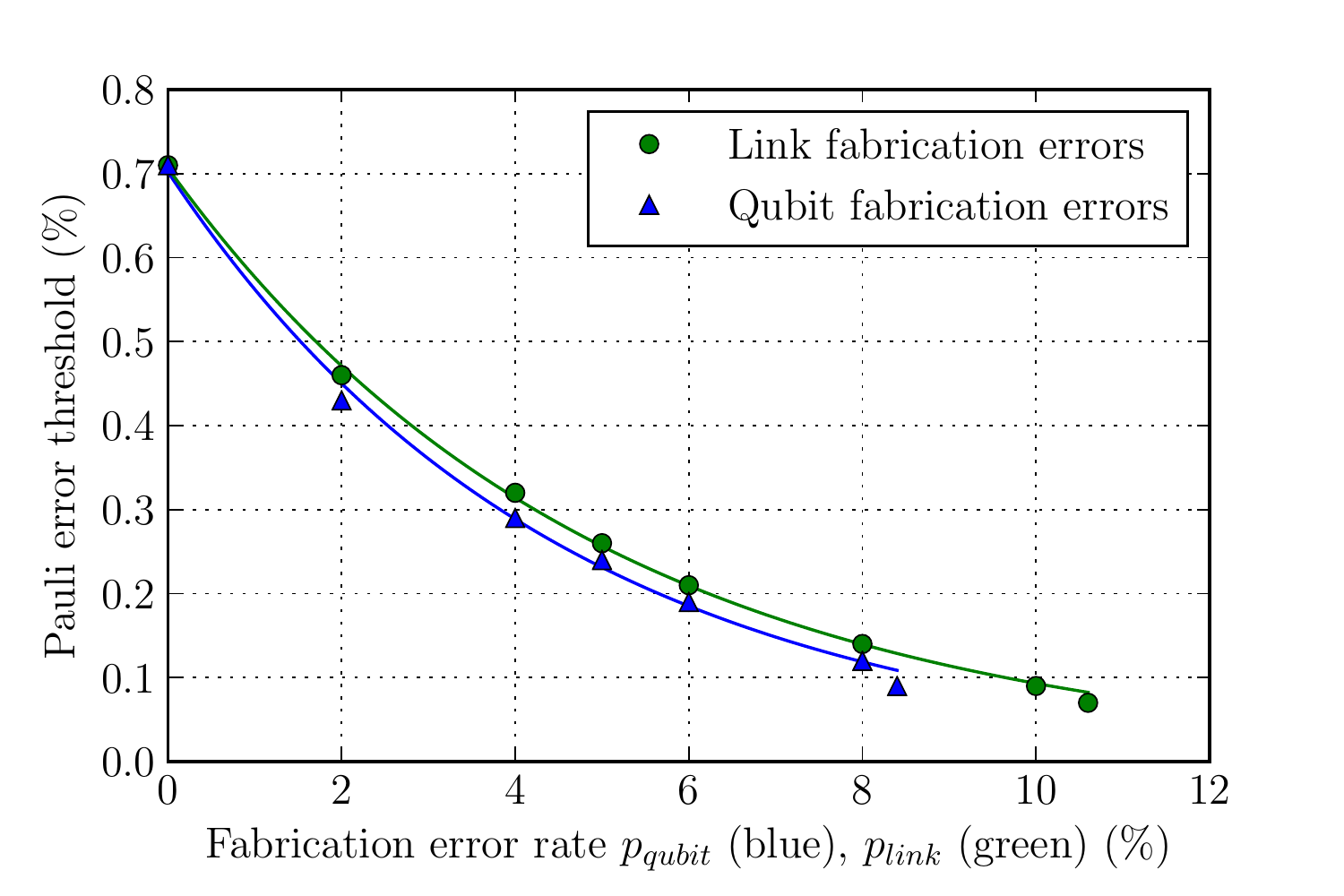}
  \caption{\label{fig:fabrication_thresholds}Pauli error thresholds for link, $p_\mathrm{link}$, and qubit, $p_\mathrm{qubit}$, fabrication errors. An exponential is fit applied to each dataset: $0.71e^{-20p_{\mathrm{link}}}$ for link fabrication errors and $0.70e^{-22p_{\mathrm{qubit}}}$ for qubit fabrication errors. Individual thresholds plots are available in the supplementary material~\cite{supplemental}.}
\end{figure}

Our main result is~\figref{fig:fabrication_thresholds}, which shows how the planar code Pauli error threshold varies with each type of fabrication error. Thresholds for each fabrication error rate are obtained by finding the crossing point of logical error rates when results for each code distance are plotted on the same graph~\cite{Wang2003-mh}; these plots are available in the supplementary material~\cite{supplemental}. Instances in which faulty or disabled data qubits percolate the lattice, such that a logical qubit cannot be encoded, are ignored, i.e. the logical error rate is given by the number of logical errors divided by the number of non-percolating runs. The percolation graphs in \figref{fig:link_perc} and \figref{fig:q_perc} give an indication of the frequency of such percolation errors.

As expected, qubit fabrication errors have a slightly more damaging effect on the threshold than link fabrication errors. With no fabrication errors, the chosen error model results in a Pauli error threshold of $p_\mathrm{comp} \approx 0.71\%$. As $p_\mathrm{qubit}$ and $p_\mathrm{link}$ increase, the thresholds of $p_\mathrm{comp}$ decrease, with the respective thresholds dropping below $0.1\%$ when $p_{qubit} \gtrsim 8\%$ and $p_{link} \gtrsim 10\%$. It has not been possible to find clean thresholds to determine the behavior of the thresholds beyond $p_\mathrm{link} = 10.6\%$ and $p_\mathrm{qubit} = 8.4\%$; this is due to effective code distances converging as fabrication error rates increase---finding thresholds requires a range of code distances, but as shown in \figref{fig:eff_dist_conn} and \figref{fig:eff_dist_qubit}, the average effective code distances for link fabrication error rates of 12\% qubit fabrication error rates of 10\% are all $\lesssim 2$ for the lattice sizes considered.

An exponential fit of the form $\alpha \exp(\beta \, p_\mathrm{fab})$ to each data set results in fits of $\alpha =0.70,\  \beta=-22$ for qubit fabrication errors and $\alpha =0.71,\  \beta=-20$ for link fabrication errors. These fits fail close to the percolation threshold but are in good agreement with the simulation results for $p_\mathrm{qubit} \le 0.08$ and $p_\mathrm{link} \le 0.1$.

\section{Discussion}

The logical error rate, $p_{log}$, depends on both the intended code distance $L$ and the actual distance $L'$, such that $p_{log} = p_{log}(L,L')$. Fabrication errors mean that $L' \le L$.  We find that $p_{log}(L',L')<p_{log}(L,L')$, i.e.~fabrication errors degrade the code performance beyond simply reducing the code distance, as shown in fig 14. The larger codes perform worse because the higher-weight supercheck operators are more prone to error and give less-specific information about the location of an error compared to a \emph{native} code. This suggests that building a larger surface code is generally only beneficial if the fabrication error rate does not increase.
\begin{figure}
  \includegraphics[width=1\linewidth]{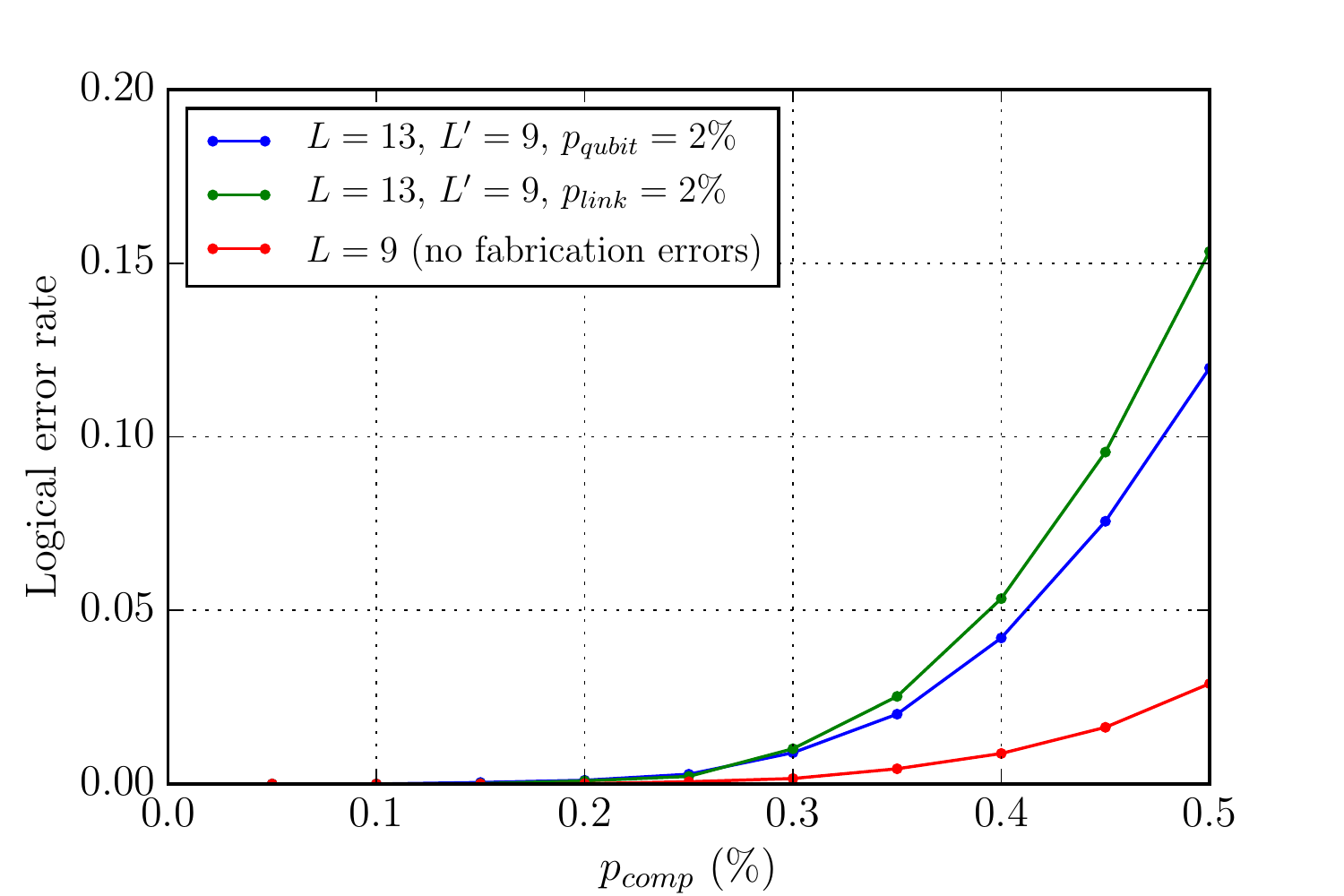}
  \caption{\label{fig:effective_vs_native_dist}Logical error rates for codes with fabrication errors with intended distance $L=13$ and actual distance $L'=9$ compared to native distance 9 codes with no fabrication errors. The native codes have lower logical error rates.}
\end{figure}

Our thresholds for qubit fabrication errors are slightly lower than that of~\cite{NFH16sce}. This is because a faulty syndrome qubit is more damaging in our scheme as each syndrome qubit fabrication error is mapped to multiple disabled data qubits; this results in lower percolation error thresholds and lower effective code distances for qubit fabrication errors. However, the gauge operator approach we present may be better for handling data qubit fabrication errors as it requires fewer logic gates. It is also worth noting that from the point of view of the implementation, our code does not require the performance of any extra gates, as the approach presented in~\cite{NFH16sce} does. All the adaptations required by our scheme involve a change in the measurement order and disabling particular measurements, not any new gates. We believe this will be a more amenable adaptation for some physical systems. Moreover, link fabrication errors were not considered in~\cite{NFH16sce}.

\begin{figure}
  \includegraphics[width=1\linewidth]{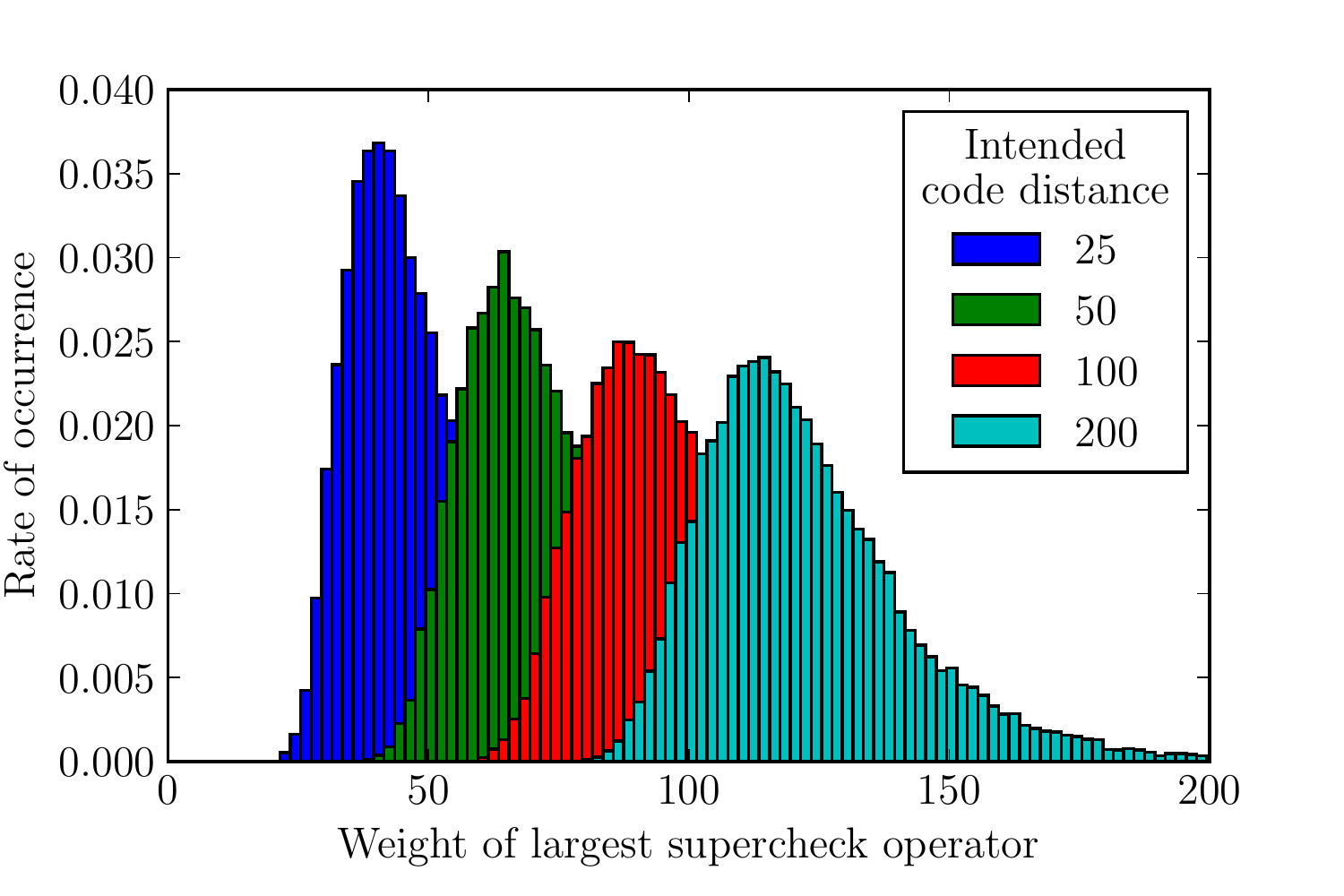}
  \caption{\label{fig:stabilizer_sizes}Rate at which highest-weight supercheck operators occur for link fabrication error rate $p_\mathrm{link} = 10\%$. The weight of the highest-weight supercheck operator appears to increase logarithmically with code distance as expected~\cite{B00lci}.}
\end{figure}

Additional simulations were performed to investigate the scaling of the weight of the largest supercheck operator as code distance increases; \figref{fig:stabilizer_sizes} shows this scaling for $p_\mathrm{link}=10\%$ and demonstrates that larger lattice sizes have higher-weight supercheck operators. This phenomenon can be understood by considering the similarity between clustering in subcritical percolation~\cite{B00lci} and forming supercheck operators on a surface code.

Higher-weight supercheck operators are more prone to error, so this effect means that the thresholds observed for smaller lattice sizes are not stable. This instability will lead to pseudo-thresholds behavior, where the threshold disappear with increasing lattice size.

This problem is not unique to our approach; it is a feature of any surface code or topological cluster state code based around the supercheck operators introduced in~\cite{SBD09tft}, including~\cite{BS10ftq} and~\cite{NFH16sce}. Finding a way to overcome this problem, or proposing an alternative to supercheck operators, remains an open problem.

\section{Conclusion}

We have presented a full analysis for the threshold performance of the surface code in the presence of fabrication errors. We showed that the ability to disable qubits or links (two-qubit gates) is sufficient to map the fabrication errors into lost qubits, such that the syndrome extraction can be performed without the need of any additional hardware components. Our method combines the supercheck operator approach and the concept of gauge qubit operators to deterministically obtain the outcome of supercheck operators. Interestingly, our approach shows that syndrome qubit fabrication errors have a more drastic damaging effect on the lattice in comparison to data qubit fabrication errors, showing that more care is required to fabricate high quality syndrome qubits. 

One advantage of the the scheme presented here is that it should be applicable to non gate-based implementations where SWAP gates may not be feasible, such as~\cite{ONR16asb}. However, in schemes where the SWAP gate is readily available,  the gauge operator scheme presented in this work and the SWAP gate scheme of~\cite{NFH16sce} can be complementary schemes. For data qubit fabrication errors, the scheme presented here requires fewer quantum operations to obtain syndrome data so would be preferable. However, the scheme in ~\cite{NFH16sce} would be preferable to deal with syndrome qubit fabrication errors, as it still allows for the measurement of stabilizer generators associated with faulty syndrome qubits. Therefore, a hybrid of both approaches could lead to an improvement in the overall performance. 

The fabrication error model presented here can be expanded to include cases where the fabrication process results in components of different quality, such that some components are fabricated with higher chance of being prone to computational Pauli errors. Such an asymmetry in the quality of fabricated components is to be expected for large-scale systems, but we leave such analysis for future investigations. 

\begin{acknowledgments}
We are grateful to Terry Rudolph and Pete Shadbolt for all the useful discussions which ultimately led to this work. We wish to thank Hector Bombin for bringing to our attention the issue of the growing size of supercheck operators with code size. JMA is supported by EPSRC. HA wishes to thank Alessio Serafini for his comments on this work and acknowledges financial support by EPSRC (grant EP/K026267/1). MGS is supported by EPSRC. TMS acknowledges funding from The Australian Research Council (ARC) Centre of Excellence in Engineered Quantum Systems, and an ARC Future Fellowship. The authors acknowledge the use of the UCL Legion High Performance Computing Facility (Legion@UCL), and associated support services, in the completion of this work. 
\end{acknowledgments}

\bibliography{bibliography.bib} 
\end{document}